\input harvmac
\input epsf
\input amssym
%\draftmode
%
%
\noblackbox
%%%%%%%%%%%%%%%%%%%%%%%%%%%%%%%%%%%%%%%%%%%%%%
%%%%%%%%%%%%%%%%%%%%%%%%%%%
% some stuff needed for figures:
%%%%%%%%%%%%%%%%%%%%%%%%%%%%%%%%%%%%%%%%%%%%%%
%%%%%%%%%%%%%%%%%%%%%%%%%%%
\newcount\figno
\figno=0
\def\fig#1#2#3{
\par\begingroup\parindent=0pt\leftskip=1cm\rightskip=1cm\parindent=0pt
\baselineskip=11pt
\global\advance\figno by 1
\midinsert
\epsfxsize=#3
\centerline{\epsfbox{#2}}
\vskip -21pt
{\bf Fig.\ \the\figno: } #1\par
\endinsert\endgroup\par
}
\def\figlabel#1{\xdef#1{\the\figno}}
\def\encadremath#1{\vbox{\hrule\hbox{\vrule\kern8pt\vbox{\kern8pt
\hbox{$\displaystyle #1$}\kern8pt}
\kern8pt\vrule}\hrule}}
%%%%%%%%%%%%%%%%%%%%%%%%%%%%%%%%%%%%%%%%%%%%%%
%%%%%%%%%%%%%%%%%%%%%%%%%%%
% definitions
%%%%%%%%%%%%%%%%%%%%%%%%%%%%%%%%%%%%%%%%%%%%%%
%%%%%%%%%%%%%%%%%%%%%%%%%%%

\def\frac#1#2{{#1 \over #2}}

\def\p{\partial}
\def\semi{\subset\kern-1em\times\;}
\def\bar#1{\overline{#1}}
\def\sqr#1#2{{\vcenter{\vbox{\hrule height.#2pt
\hbox{\vrule width.#2pt height#1pt \kern#1pt \vrule width.#2pt}
\hrule height.#2pt}}}}

\def\p{\partial}

\def\th{\theta}

\def\ad{\bar a}

\def\th{\hat{t}}

\def\p{\partial}

%
%%%%%%%%%%%%%%%%%%%%%%%%%%%%%%%%%%%%%%%%%%%%%%
%%%%%%%%%%%%%%%%%%%%%%%%%%%
% more definitions
%%%%%%%%%%%%%%%%%%%%%%%%%%%%%%%%%%%%%%%%%%%%%%
%%%%%%%%%%%%%%%%%%%%%%%%%%%

%
%\def\oneone{\rlap 1\mkern4mu{\rm l}}
%\def\coeff#1#2{\relax{\textstyle {#1 \over #2}}\displaystyle}

\def\IR{\Bbb{R}}

\def\cV{ {\cal V}}
\def\cA{ {\cal A}}
\def\zb{\overline{z}}
\def\th{\hat{t}}

%%%%%%%%%%%%%%%%%%%%%%%%%%%%%%%%%%%%%%%%%%%%%%
%%%%%%%%%%%%%%%%%%%%%%%%%%%
% References
%%%%%%%%%%%%%%%%%%%%%%%%%%%%%%%%%%%%%%%%%%%%%%
%%%%%%%%%%%%%%%%%%%%%%%%%%%

\lref\attract{
S.~Ferrara, R.~Kallosh and A.~Strominger,
``N=2 extremal black holes'',
Phys.\ Rev.\ D {\bf 52}, 5412 (1995),
[arXiv:hep-th/9508072];
  %%CITATION = HEP-TH 9508072;%%
 A.~Strominger,
 ``Macroscopic Entropy of $N=2$ Extremal Black Holes'',
 Phys.\ Lett.\ B {\bf 383}, 39 (1996),
[arXiv:hep-th/9602111];
  %%CITATION = HEP-TH 9602111;%%
S.~Ferrara and R.~Kallosh,
``Supersymmetry and Attractors'',
Phys.\ Rev.\ D {\bf 54}, 1514 (1996),
[arXiv:hep-th/9602136];
  %%CITATION = HEP-TH 9602136;%%
``Universality of Supersymmetric Attractors'',
Phys.\ Rev.\ D {\bf 54}, 1525 (1996),
[arXiv:hep-th/9603090];
 %%CITATION = HEP-TH 9603090;%%
R.~Kallosh, A.~Rajaraman and W.~K.~Wong,
``Supersymmetric rotating black holes and attractors'',
Phys.\ Rev.\ D {\bf 55}, 3246 (1997),
[arXiv:hep-th/9611094];
  %%CITATION = HEP-TH 9611094;%%
A~Chou, R.~Kallosh, J.~Rahmfeld, S.~J.~Rey, M.~Shmakova and W.~K.~Wong,
``Critical points and phase transitions in 5d compactifications of M-theory''.
Nucl.\ Phys.\ B {\bf 508}, 147 (1997);
[arXiv:hep-th/9704142].
 %%CITATION = HEP-TH 9704142;%%
}

\lref\moore{G.~W.~Moore,``Attractors and arithmetic'',
[arXiv:hep-th/9807056];
%%CITATION = HEP-TH 9807056;%%
``Arithmetic and attractors'',
[arXiv:hep-th/9807087];
  %%CITATION = HEP-TH 9807087;%%
``Les Houches lectures on strings and arithmetic'',
[arXiv:hep-th/0401049];
  %%CITATION = HEP-TH 0401049;%%
B.~R.~Greene and C.~I.~Lazaroiu,
``Collapsing D-branes in Calabi-Yau moduli space. I'',
Nucl.\ Phys.\ B {\bf 604}, 181 (2001),
[arXiv:hep-th/0001025].
}

%\ChamseddinePI
\lref\ChamseddinePI{
  A.~H.~Chamseddine, S.~Ferrara, G.~W.~Gibbons and R.~Kallosh,
  ``Enhancement of supersymmetry near 5d black hole horizon,''
  Phys.\ Rev.\ D {\bf 55}, 3647 (1997)
  [arXiv:hep-th/9610155].
  %%CITATION = HEP-TH 9610155;%%
}

\lref\denef{  %%CITATION = HEP-TH 0001025;%%
F.~Denef,``Supergravity flows and D-brane stability'',
JHEP {\bf 0008}, 050 (2000), [arXiv:hep-th/0005049];
%%CITATION = HEP-TH 0005049;%%
``On the correspondence between D-branes and stationary supergravity
 solutions of type II Calabi-Yau compactifications'',
[arXiv:hep-th/0010222];
 %%CITATION = HEP-TH 0010222;%%
``(Dis)assembling special Lagrangians'',
[arXiv:hep-th/0107152].
%%CITATION = HEP-TH 0107152;%%
  B.~Bates and F.~Denef,
   ``Exact solutions for supersymmetric stationary black hole composites,''
  arXiv:hep-th/0304094.
}

\lref\topstr{H.~Ooguri, A.~Strominger and C.~Vafa,
``Black hole attractors and the topological string'',
Phys.\ Rev.\ D {\bf 70}, 106007 (2004),
[arXiv:hep-th/0405146];
 %%CITATION = HEP-TH 0405146;%%
E.~Verlinde,
``Attractors and the holomorphic anomaly'',
[arXiv:hep-th/0412139];
  %%CITATION = HEP-TH 0412139;%%
A.~Dabholkar, F.~Denef, G.~W.~Moore and B.~Pioline,
``Exact and asymptotic degeneracies of small black holes'',
[arXiv:hep-th/0502157].
  %%CITATION = HEP-TH 0502157;%%
}

\lref\curvcorr{A.~Dabholkar,
``Exact counting of black hole microstates",
[arXiv:hep-th/0409148],
%%CITATION = HEP-TH 0409148;%%
A.~Dabholkar, R.~Kallosh and A.~Maloney,
``A stringy cloak for a classical singularity'',
JHEP {\bf 0412}, 059 (2004),
[arXiv:hep-th/0410076].
 %%CITATION = HEP-TH 0410076;%%
}
\lref\bkmicro{
 I.~Bena and P.~Kraus,
 ``Microscopic description of black rings in AdS/CFT'',
JHEP {\bf 0412}, 070 (2004)
  [arXiv:hep-th/0408186].
%%CITATION = HEP-TH 0408186;%%
}
\lref\cgms{
M.~Cyrier, M.~Guica, D.~Mateos and A.~Strominger,
``Microscopic entropy of the black ring'',
[arXiv:hep-th/0411187].
  %%CITATION = HEP-TH 0411187;%%
}

%\GunaydinBI
\lref\GunaydinBI{
  M.~Gunaydin, G.~Sierra and P.~K.~Townsend,
  ``The Geometry Of N=2 Maxwell-Einstein Supergravity And Jordan
  Algebras'',
  Nucl.\ Phys.\ B {\bf 242}, 244 (1984);
  ``Gauging The D = 5 Maxwell-Einstein Supergravity Theories:
   More On Jordan Algebras,''
  Nucl.\ Phys.\ B {\bf 253}, 573 (1985).
  %%CITATION = NUPHA,B253,573;%%
}

%\deWitCR
\lref\deWitCR{
  B.~de Wit and A.~Van Proeyen,
  ``Broken sigma model isometries in very special geometry,''
  Phys.\ Lett.\ B {\bf 293}, 94 (1992)
  [arXiv:hep-th/9207091].
  %%CITATION = HEP-TH 9207091;%%
}

%\CadavidBK
\lref\CadavidBK{
  A.~C.~Cadavid, A.~Ceresole, R.~D'Auria and S.~Ferrara,
  ``Eleven-dimensional supergravity compactified on Calabi-Yau threefolds,''
  Phys.\ Lett.\ B {\bf 357}, 76 (1995)
  [arXiv:hep-th/9506144].
  %%CITATION = HEP-TH 9506144;%%
}

%\PapadopoulosDA
\lref\PapadopoulosDA{
  G.~Papadopoulos and P.~K.~Townsend,
  ``Compactification of D = 11 supergravity on spaces of exceptional
  holonomy,''
  Phys.\ Lett.\ B {\bf 357}, 300 (1995)
  [arXiv:hep-th/9506150].
  %%CITATION = HEP-TH 9506150;%%
}

%\AntoniadisCY
\lref\AntoniadisCY{
  I.~Antoniadis, S.~Ferrara and T.~R.~Taylor,
  ``N=2 Heterotic Superstring and its Dual Theory in Five Dimensions,''
  Nucl.\ Phys.\ B {\bf 460}, 489 (1996)
  [arXiv:hep-th/9511108].
  %%CITATION = HEP-TH 9511108;%%
}

%\GauntlettNW
\lref\GauntlettNW{
  J.~P.~Gauntlett, J.~B.~Gutowski, C.~M.~Hull, S.~Pakis and H.~S.~Reall,
  ``All supersymmetric solutions of minimal supergravity in five dimensions,''
  Class.\ Quant.\ Grav.\  {\bf 20}, 4587 (2003)
  [arXiv:hep-th/0209114]}
  %%CITATION = HEP-TH 0209114;%%}

\lref\gutow{
  J.~B.~Gutowski and H.~S.~Reall,
  ``General supersymmetric AdS(5) black holes'',
  JHEP {\bf 0404}, 048 (2004)
  [arXiv:hep-th/0401129];
  J.~B.~Gutowski,
  ``Uniqueness of five-dimensional supersymmetric black holes'',
  JHEP {\bf 0408}, 049 (2004)
  [arXiv:hep-th/0404079].
}

%\BenaDE
\lref\BenaDE{
  I.~Bena and N.~P.~Warner,
  ``One ring to rule them all ... and in the darkness bind them?'',
  [arXiv:hep-th/0408106].
  %%CITATION = HEP-TH 0408106;%%
}
%\BMPV
\lref\BMPV{ J.~C.~Breckenridge, R.~C.~Myers, A.~W.~Peet and C.~Vafa,
``D-branes and spinning black holes'',
Phys.\ Lett.\ B {\bf 391}, 93 (1997);
[arXiv:hep-th/9602065].
  %%CITATION = HEP-TH 9602065;%%
}

\lref\EEMR{H.~Elvang, R.~Emparan, D.~Mateos and H.~S.~Reall,
``Supersymmetric black rings and three-charge supertubes'',
  Phys.\ Rev.\ D {\bf 71}, 024033 (2005);
  [arXiv:hep-th/0408120].
  %%CITATION = HEP-TH 0408120;%%
}

%\ElvangRT
\lref\ElvangRT{
  H.~Elvang, R.~Emparan, D.~Mateos and H.~S.~Reall,
  ``A supersymmetric black ring'',
  Phys.\ Rev.\ Lett.\  {\bf 93}, 211302 (2004)
  [arXiv:hep-th/0407065].
  %%CITATION = HEP-TH 0407065;%%
}

%\BenaWT
\lref\BenaWT{
  I.~Bena and P.~Kraus,
  ``Three charge supertubes and black hole hair,''
  Phys.\ Rev.\ D {\bf 70}, 046003 (2004)
  [arXiv:hep-th/0402144].
  %%CITATION = HEP-TH 0402144;%%
}

%\GauntlettQY
\lref\GauntlettQY{ J.~P.~Gauntlett and J.~B.~Gutowski, ``General
Concentric  Black Rings'', [arXiv:hep-th/0408122]. J.~P.~Gauntlett
and J.~B.~Gutowski, ``Concentric  black rings'',
[arXiv:hep-th/0408010].
%%CITATION = HEP-TH 0408122;%%
}

%\DenefNB
\lref\denefa{
  F.~Denef,
   ``Supergravity flows and D-brane stability'',
JHEP {\bf 0008}, 050 (2000), [arXiv:hep-th/0005049].
  %%CITATION = HEP-TH 0005049;%%
}

%\BatesVX
\lref\denefc{
  B.~Bates and F.~Denef,
``Exact solutions for supersymmetric stationary black hole composites'',
[arXiv:hep-th/0304094].
  %%CITATION = HEP-TH 0304094;%%
}

%\SenPU
\lref\SenPU{
A.~Sen,
``Black holes, elementary strings and holomorphic anomaly'',
 [arXiv:hep-th/0502126].
%%CITATION = HEP-TH 0502126;%%
}

%\CardosoFP
\lref\CardosoFP{
G.~L.~Cardoso, B.~de Wit, J.~Kappeli and T.~Mohaupt,
``Examples of stationary BPS solutions in N = 2 supergravity theories  with
$R^2$-interactions,''
Fortsch.\ Phys.\  {\bf 49}, 557 (2001)
[arXiv:hep-th/0012232];
``Stationary BPS solutions in N = 2 supergravity with $R^2 $ interactions'',
JHEP {\bf 0012}, 019 (2000)
[arXiv:hep-th/0009234];
  ``Supersymmetric black hole solutions with $R^2$ interactions'',
[arXiv:hep-th/0003157];
  G.~Lopes Cardoso, B.~de Wit and T.~Mohaupt,
  ``Area law corrections from state counting and supergravity'',
  Class.\ Quant.\ Grav.\  {\bf 17}, 1007 (2000)
  [arXiv:hep-th/9910179];
  ``Macroscopic entropy formulae and non-holomorphic corrections for
  supersymmetric black holes'',
  Nucl.\ Phys.\ B {\bf 567}, 87 (2000)
  [arXiv:hep-th/9906094];
  ``Deviations from the area law for supersymmetric black holes'',
  Fortsch.\ Phys.\  {\bf 48}, 49 (2000)
  [arXiv:hep-th/9904005];
  ``Corrections to macroscopic supersymmetric black-hole entropy'',
  Phys.\ Lett.\ B {\bf 451}, 309 (1999)
  [arXiv:hep-th/9812082].
  %%CITATION = HEP-TH 9812082;%%
}

%\BenaTD
\lref\BenaTD{
  I.~Bena, C.~W.~Wang and N.~P.~Warner,
  ``Black rings with varying charge density'',
[arXiv:hep-th/0411072].
  %%CITATION = HEP-TH 0411072;%%
}

%\BenaWV
\lref\BenaWV{
I.~Bena, ``Splitting hairs of the three charge black hole'',
Phys.\ Rev.\ D {\bf 70}, 105018 (2004),
[arXiv:hep-th/0404073].
  %%CITATION = HEP-TH 0404073;%%
}

%%%%%%%%%%%%%%%%%%%%%%%%%%%%%%%%%%%%%%%%%%%%%%
%%%%%%%%%%%%%%%%%%%%%%%%%%%
% Title
%%%%%%%%%%%%%%%%%%%%%%%%%%%%%%%%%%%%%%%%%%%%%%
%%%%%%%%%%%%%%%%%%%%%%%%%%%

\Title{\vbox{\baselineskip12pt \hbox{hep-th/0503219}
\hbox{UCLA-05-TEP-09} \hbox{MCTP-05-68}}} {\vbox{\centerline
{Attractors and Black Rings}} } \centerline{Per
Kraus\foot{pkraus@physics.ucla.edu} and Finn
Larsen\foot{larsenf@umich.edu}}

\bigskip
\centerline{${}^1$\it{Department of Physics and Astronomy,
UCLA,}}\centerline{\it{ Los Angeles, CA 90095-1547,
USA.}}\vskip.2cm \centerline{${}^2$\it{Michigan Center for
Theoretical Physics,}} \centerline{\it{University of Michigan, Ann
Arbor, MI 48109-1120, USA.}}

\baselineskip15pt

\vskip .3in

\centerline{\bf Abstract} {The attractor mechanism is usually
thought of as the fixing of the near horizon moduli of a BPS black
hole in terms of conserved charges measured at infinity.   Recent
progress in understanding BPS solutions in five dimensions
indicates that this is an incomplete story.  Moduli can instead be
fixed in terms of dipole charges, and their corresponding values
can be found  by extremizing a certain attractor function built
out of these charges. BPS black rings provide an example of this
phenomenon.  We give a general derivation of the attractor
mechanism in five dimensions based on the recently developed
classification of BPS solutions. This analysis shows when it is
the  dipole charges versus the conserved charges that fix the
moduli. It also yields explicit expressions for the fixed moduli.
}

%%%
\Date{March, 2005}
%%%%%%%%%%%%%%%%%%%%%%%%%%%%%%%%%%%%%%%%%%%%%%
%%%%%%%%%%%%%%%%%%%%%%%%%%%
% Main text begins here
%%%%%%%%%%%%%%%%%%%%%%%%%%%%%%%%%%%%%%%%%%%%%%
%%%%%%%%%%%%%%%%%%%%%%%%%%%
\baselineskip14pt
\newsec{Introduction}

Our ability to make microscopic sense of the entropy of BPS black
holes rests on, among other things, the attractor mechanism: the
property that the scalar moduli fields at the horizon are fixed in
terms of charges carried by the black hole.\foot{More precisely,
it is the vectormultiplet moduli which are fixed.  The
hypermultiplet moduli are not fixed, but do not affect the black
hole entropy.} The point is that the moduli are continuous
parameters which can be freely specified at infinity, raising the
dangerous possibility that the entropy might depend on their
values. Such a dependence  presumably would lead to a violation of
the second law of thermodynamics, since it would allow one to
quasi-statically decrease the entropy by varying the moduli. What
saves the day is that the entropy depends only on the values of
the moduli at the horizon, and these turn out to be insensitive to
the values at infinity.  The black hole entropy thus ends being a
function purely of the charges.

The existing literature on the attractor mechanism deals with
spherically symmetric  black holes in four dimensions, and
spherically symmetric or rotating BMPV black holes in five
dimensions
\refs{\attract,\moore,\ChamseddinePI}.\foot{Multi-centered black
holes in four dimensions have also been considered, and lead to
some of the same issues discussed in this paper \denef.}  For
these examples the attractor mechanism works in a simple way.  The
black holes carry conserved electric and magnetic charges, which
can be measured at infinity in terms of flux integrals.  The BPS
mass formula is a particular combination of these charges and the
values of the asymptotic moduli.  If one computes the values of
the moduli which minimize the BPS mass as a function of the
charges, then it turns out that these are the same values as
obtained by the moduli at the horizon.  The derivation of this
result uses the fact that the amount of supersymmetry preserved by
the black hole solution is enhanced near the horizon.  Vanishing
of the corresponding Killing spinor equations leads to contraints
on the moduli.

We now appreciate that the above class of BPS black hole solutions
is far from the complete story.  Recent work has provided a much
better understanding of the  structure of the BPS equations
governing general solutions \refs{\GauntlettNW,\gutow}, and has
led to interesting new examples, such as black rings in five
dimensions
\refs{\ElvangRT,\BenaDE,\EEMR,\GauntlettQY,
\BenaTD,\BenaWT,\BenaWV,\bkmicro,\cgms}.
Does the attractor mechanism function in this general context and,
if so, how?  This is the question that we address here.

In fact, a quick perusal of the black ring solution makes it
evident that the attractor mechanism is functioning in a different
way.  First of all, the black ring entropy is not purely a
function of the conserved charges, but also depends on the values
of {\it dipole charges}, which are non-conserved quantities
measured by flux integrals on surfaces linked with the ring.  From
this point of view it would not be a surprise if it turns out that
the moduli at the horizon also depend on the dipole charges.
Indeed, the moduli turn out to be determined {\it entirely} by the
dipole charges.  Obviously, the corresponding values cannot be
ascertained by extremizing the BPS mass as above, since the BPS
mass depends only on the conserved charges and not the dipole
charges.

With this in mind, we will carefully rederive the details of the
attractor mechanism in a general context.  We will work in the
five dimensional $N=2$ supergravity corresponding to
compactification of M-theory on a Calabi-Yau threefold.  The five
dimensional setting is advantageous since it leads to simpler
formulas, and it is also the habitat of the black ring.  It would
of course be interesting to extend our results to the four
dimensional context, noting that not every four dimensional case
can be obtained via dimensional reduction from five dimensions.
The four dimensional case has also been the subject of several
interesting recent developments
\refs{\CardosoFP,\topstr,\curvcorr,\SenPU}.

We will strive to be as general as possible, going as far as we
can based just on the general BPS equations obeyed by any BPS
solution.  It will become apparent that there are two distinct
cases to consider, depending on whether or not certain components
of the field strengths (the {\it dipole field strengths}) are zero
or nonzero.   When they vanish, which is the case for the BMPV
black hole, we will reproduce earlier results showing that the
moduli are fixed in terms of the conserved charges.  This can be
phrased in terms of extremizing the central charge $Z_e$, defined
as
\eqn\inta{ Z_e = X^I Q_I~,}
where $X^I$ are the vectormultiplet moduli.   We also demonstrate
that the flow of $Z_e$ from infinity to the horizon is
monotonically decreasing, in parallel to what is known for
spherically symmetric solutions in four dimensions.  At infinity
$Z_e$ gives the BPS mass, while at the horizon it fixes the
central charge of the CFT dual to the near-horizon AdS geometry.

For nonvanishing dipole field strengths the story changes in an
essential way;  for instance $Z_e$ no longer behaves
monotonically, and indeed typically diverges at the horizon.
Instead, a new attractor function $Z_m$ takes over, defined as
\eqn\intb{ Z_m = X_I q^I~,}
where $q^I$ are the dipole charges.  For sufficiently many nonzero
$q^I$, extremization of  $Z_m$ yields the near horizon values of
the moduli.  Further, the value of $Z_m$ at the horizon determines
the central charge of the associated CFT.

In the case of the black ring solution, it turns out that $Z_m$ is
proportional to a certain combination of the angular momenta, $Z_m
\propto J_\psi - J_\phi$, and so the near-horizon moduli can
equivalently be determined by extremizing this quantity.  This is
analogous to extremizing the BPS mass in the case of the BMPV
class of solutions. Extremization of $Z_m$ also makes sense from
another point of view.  If the ring direction of the black ring
was instead an infinite line, then we would have a magnetic string
solution whose BPS mass is proportional to $Z_m$.

The remainder of this paper is organized as follows.  In section 2
we review the relevant aspects of real special geometry, which is
the appropriate language for five dimensional $N=2$ supergravity.  The
constraints of supersymmetry are reviewed in section 3.  Section
4, which is the core of the paper, gives the general derivation of
the attractor mechanism.  We close with a brief discussion in section 5.

\newsec{Review of Real Special Geometry}

The general setting for our study is the five dimensional low
energy  supergravity theory corresponding to M-theory compactified
on a Calabi-Yau threefold CY$_3$.  This subject goes by the name
{\it real special geometry} or {\it very special geometry}.  It is
in several ways simpler than the special geometry employed in
compactifications to four dimensions, albeit perhaps less
familiar. Relevant references include
\refs{\GunaydinBI,\deWitCR,\CadavidBK,\PapadopoulosDA,\AntoniadisCY,
\ChamseddinePI}. Our notation will mainly follow \ChamseddinePI.
In particular, we use a mostly plus signature metric, and the
Clifford algebra reads $\left\{ \Gamma^\mu, \Gamma^\nu \right\} =
2g^{\mu\nu}$. The unit of length is the $D=11$ Planck length
$l_p$, which we set equal to $1$:   $l_p = (\pi/4G_5)^{1/3}=1$.

We take our CY$_3$ to have Hodge numbers $h^{1,1}$ and $h^{2,1}$.
Let $J_I$ be a basis of $(1,1)$ forms, with $I=1,2, \ldots ,
h^{1,1}$, and expand the K\"{a}hler form $J$ on the CY$_3$ as
\eqn\aa{ J = X^I J_I~.}
This defines the K\"{a}hler moduli $X^I$ which are {\it real}
\foot{The notation $t^I \equiv 6^{-1/3}X^I$ is common in the
literature, including \ChamseddinePI.}. In terms of homology the
K\"{a}hler moduli correspond to the volumes of the 2-cycles
$\Omega^I$
\eqn\aab{ X^I = \int_{\Omega^I} J~.}
The triple intersection numbers are defined as
\eqn\aac{C_{IJK} = \int_{CY} J_I \wedge J_J \wedge J_K~.}
The terminology arises because this quantity can equally well be
defined in terms of homology and then the integral just counts the
intersection points of three 4-cycles $\Omega_I$, $\Omega_J$, and
$\Omega_K$. The volumes of the 4-cycles $\Omega_I$ are given
by\foot{Our convention for $X_I$ differ from some papers ({\it
e.g.} \ChamseddinePI ) by $X_I^{\rm here} = 3X_I^{\rm there}$.}
\eqn\aaa{X_I = {1\over 2}\int_{\Omega_I} J\wedge J=  {1\over
2}\int_{CY} J\wedge J\wedge J_I = {1\over 2} C_{IJK} X^J X^K~.}

The K\"{a}hler moduli are the lower components of $N=2$
vectormultiplets. The couplings of these are entirely determined
by the prepotential
\eqn\ab{ \cV = {1\over 3!} \int_{CY} J\wedge J \wedge J  = {1\over
6} C_{IJK} X^I X^J X^K~.}
This expression is interpreted geometrically as the overall
volume of the CY$_3$ and is a component of a hypermultiplet.  In
this paper we will be ignoring the hypermultiplets in the sense
that we consistently set them to fixed constant values.
Therefore, we impose the condition
\eqn\ac{ \cV = {1\over 6} C_{IJK} X^I X^J X^K =1~.}
This is to be understood as a constraint on the K\"{a}hler moduli,
to be imposed after varying the prepotential to derive the
equations  of motion. There are thus $n_v= h^{1,1}-1$ independent
vectormultiplets. We denote the $n_v$ independent vectormultiplet
scalars as $\phi^i$, and the corresponding derivatives $\p_i = {\p
\over \p \phi^i}$.

Reduction of the 3-form potential in M-theory gives rise to $h^{1,1}$
1-form potentials $A^I$ in $D=5$:
\eqn\ad{\cA = A^I \wedge J_I~.}
The linear combination $X_I A^I$ is identified as the  graviphoton
in the gravity multiplet. The remaining $n_v$ gauge fields are the
upper components of the $N=2$ vectormultiplets. The kinetic terms
for the gauge fields are governed by the metric
\eqn\ag{\eqalign{G_{IJ} = \half\int_{CY} J_I\wedge {}^\star J_J =
-\half (\p_I \p_J \ln \cV)_{\cV=1} &=-\half( C_{IJK} X^K -  X_I
X_J)~,}}
where we use the notation for derivatives:  $\p_I = {\p \over \p
X^I}$.  The bosonic part of the $D=5$ action is
\eqn\ai{ {\cal L^{(\rm bos)}}  ={1 \over 16\pi G_5} \left\{
-R~ {}^*1-  G_{IJ}  dX^I  \wedge {}^*dX^J  - G_{IJ}
F^{I} \wedge {}^*F^{J} - {1\over 6}
 C_{IJK} F^I \wedge
F^J \wedge A^K\right\}~.}

Let us note some useful relations. First, \aaa\  and \ac\  give
\eqn\aca{X_I X^I =3~,}
and so
\eqn\aib{ X^I \p_i X_I  = \p_i X^I X_I =0~.}
Next, combining these relations with \ag\ we find
\eqn\aic{\quad X_I = 2 G_{IJ}X^J~, \quad \p_i X_I= - 2 G_{IJ} \p_i
X^J~.} Using the metric $2G_{IJ}$ to lower the indices
$I,J,\cdots$ we can introduce the 2-cycle intersection numbers
$C^{IJK}$ normalized such that the volume condition on the 3-fold
becomes ${1 \over 6 } C^{IJK} X_I X_J X_K =1$. (Since 2-cycles do
not in general intersect, the geometric interpretation of these
numbers refers to the intersection of  the dual 4-cycles.)

We will often consider the simplest case of $T^6$, in which case
we write the metric and 3-form as
\eqn\aj{\eqalign{ ds^2 & = ds_5^2 + X^1 dz_1 d\zb_1 + X^2 dz_2
d\zb_2 + X^3 dz_3 d\zb_3 \cr \cA & = A^1 \wedge ({i \over 2} dz_1
\wedge d\zb_1) +A^2 \wedge ({i \over 2} dz_2 \wedge d\zb_2) +A^3
\wedge ({i \over 2} dz_3 \wedge d\zb_3)~.}}
In this case $C_{IJK}=1$ if $(IJK)$ is a permutation of $(123)$,
and $C_{IJK}=0$ otherwise.   The metric $G_{IJ}$ is
\eqn\ak{ G_{IJ}= \half {\rm diag} \left(
(X^1)^{-2},(X^2)^{-2},(X^3)^{-2}\right)~.}
We also have the relations
\eqn\al{X^1 X^2 X^3=1~, \quad X_I = {1 \over X^I}~.}

\newsec{BPS Equations}

We will be interested in solutions preserving some supersymmetry.
For purely bosonic backgrounds we need to set to zero the
supersymmetry variations of the gravitinos and the gauginos
\eqn\ba{\eqalign{ \delta \psi_\mu &= \left[ D_\mu(\omega)  +  {i
\over 24}X_I (\Gamma_\mu^{~\nu\rho} -4 \delta_\mu^\nu \Gamma^\rho)
F^I_{\nu\rho} \right]\epsilon~, \cr \delta \lambda_i &= -{1 \over
4} G_{IJ} \p_i X^{I} F^J_{\mu\nu} \Gamma^{\mu\nu} \epsilon -
{i\over 2} G_{IJ}\p_i X^I \Gamma^\mu \p_\mu X^J \epsilon~. }}
Here  $\Gamma^{\mu\nu} = {1 \over 2}(\Gamma^\mu \Gamma^\nu -
\Gamma^\nu \Gamma^\mu)$.

Preservation of some supersymmetry implies conditions on the
bosonic fields, and these have been massaged into a compact and
useable form in \GauntlettNW\gutow. Supersymmetry implies the
existence of a Killing vector, and assuming that it  is time-like
we first write the $D=5$ metric in the form
\eqn\bb{ ds_5^2 = -f^2 (dt+\omega)^2 +f^{-1} h_{mn} dx^m dx^n~,}
where $h_{mn}$ is a hyper-K\"{a}hler metric on the {\it base} ${\cal
B}$, and $\omega$ is a 1-form on ${\cal B}$.  We then define the
{\it dipole field strengths} $\Theta^I$ by writing the field
strengths $F^I = dA^I$ in the form
\eqn\bc{F^I = d\left[(fX^I(dt+\omega)\right] +\Theta^I~,}
where the $\Theta^I$ are closed 2-forms on the base. The BPS equations
then take the form\foot{This form presumably remains  valid after the
assumption of a symmetric scalar manifold \gutow\ is relaxed.}
\eqn\bd{ \eqalign{ \Theta^I &= {}^{\star_4} \Theta^I \cr  \nabla^2
(f^{-1} X_I)&= {1 \over 4} C_{IJK} \Theta^J \cdot \Theta^K \cr
d\omega + {}^{\star_4} d\omega & = - f^{-1} X_I \Theta^I~,}}
where $\nabla^2$  and ${}^{\star_4}$ are defined with respect to the
base ${\cal B}$, and for 2-forms $\alpha$ and $\beta$ on ${\cal
B}$ we define $\alpha \cdot \beta = \alpha^{mn } \beta_{mn}$, with
indices raised by $h^{mn}$. The BPS equations written in the above
order are linear, as noted in \BenaDE.

\newsec{Attractor Mechanism}

\subsec{The Flow Equation}
The attractor equations ultimately follow from the gaugino variation
in \ba. In this subsection we derive an important intermediate result,
a flow equation relating the flow of the moduli to changes
in the gauge field.

We look for solutions of $\delta \lambda_i=0$ with the spinor
$\epsilon$ obeying
\eqn\ca{ \Gamma^{\th} \epsilon = -i \epsilon~.}
Hatted indices are with respect to an orthonormal frame.  In order
that we preserve half the supersymmetries, we demand that
$\epsilon$ be subject to no other projection equations.  From \bc\
the components of the field strength are
\eqn\cb{ \eqalign{F^I_{m \th} &= f^{-1}  \p_m(fX^I)~,\cr
 F^I_{mn} & = f X^I
(d\omega)_{mn}+ \Theta^I_{mn}~.}}
The purely spatial components of $F^I$,   displayed in the second
line of \cb, do not contribute to the gaugino variation equations:
the first term can be eliminated using \aib; and the self-duality
of $\Theta^I$ and the projection \ca\ shows that the second term
does not contribute either. The contribution of the remaining
components $F^I_{m\th}$ to the gaugino variation implies the
equation
\eqn\cc{ G_{IJ} \p_i X^I F^J_{m\th} = G_{IJ} \p_i X^I \p_m X^J~.}
We now proceed to manipulate \cc.  Define the electric
field as
\eqn\cd{ E_{m I} = G_{IJ} F^J_{m\th}~,}
and take the following divergence with respect to the metric
$h_{mn}$:
\eqn\ce{\eqalign{ \nabla^m (f^{-1} X^I E_{mI}) &=(\nabla^m X^I
)f^{-1} E_{mI} + X^I \nabla^m (f^{-1} E_{mI}) \cr &= f^{-1} G_{IJ}
\nabla^m X^I \nabla_m X^J +X^I \nabla^m (f^{-1} E_{mI})~,}}
where we used \cc\ to arrive at the second line.  Next, we write
\eqn\cfa{\eqalign{  \nabla^m [f^{-1}E_{mI}] &= \nabla^m [ f^{-2}
G_{IJ} \partial_m (fX^J)] \cr &=  {1\over 2} \nabla^m [ f^{-2}
(\partial_m f) X_I - f^{-1} \partial_m X_I ] \cr &=  - {1\over 2}
\nabla^m\partial_m [ f^{-1}X_I]\cr &= -{1\over 8} C_{IJK} \Theta^J
\cdot \Theta^K~,}}
using first the definitions \cb\ and \cd, then both equations in \aic,
and finally the BPS equation \bd. Inserting back in to \ce\ we obtain
\eqn\ch{ \nabla^m (f^{-1} X^I E_{mI}) = f^{-1} G_{IJ} \nabla^m X^I
\nabla_m X^J  - {1 \over 8} C_{IJK}X^I \Theta^J \cdot \Theta^K~.}
This is the flow equation we wanted to derive.

\subsec{Near Horizon Enhancement of SUSY}
BPS black hole solutions typically exhibit a type of domain wall
structure, interpolating between two maximally supersymmetric
vacua of the theory: Minkowski space at infinity, and AdS at the
horizon.  From the point of view of gauge/gravity duality, this is
interpreted as the renormalization group flow to an infrared fixed
point CFT.  The phenomenon of enhanced supersymmetry leads
to strong constraints on the values of the vectormultiplet moduli
at the horizon, as we now discuss.

We return to the supersymmetry variations \ba, and now demand that
we can set $\delta \psi_\mu = \delta \lambda_i=0$ without imposing
any projection conditions on $\epsilon$ analogous to \ca.
Vanishing of the gravitino variation implies that the metric is
maximally symmetric, {\it e.g.} AdS$_2\times S^3$ or AdS$_3 \times
S^2$; see \ChamseddinePI\ for details.  We focus instead on the
gaugino variation.   For general $\epsilon$ there is no
possibility of a cancellation among the terms in $\delta
\lambda_i$, and so we need to impose
\eqn\ci{\eqalign{G_{IJ}\p_i X^I \p_\mu X^J &=0~, \cr G_{IJ} \p_i
X^{I} F^J_{\mu\nu} =0~, \quad &{\rm or~equivalently}\quad \p_i X_I
F^I_{\mu\nu}=0~.}}
Multiplying the first equation by $\p_\mu \phi^i$ and contracting
the $\mu$ indices gives
\eqn\cia{G_{IJ}g^{\mu\nu}  \p_\mu X^I \p_\nu X^J
=0~.} For a static configuration, this is positive semi-definite,
and so implies constant moduli:
\eqn\cj{ \p_\mu X^I =0~.}
Then, assuming constant moduli, the field strength components \cb\
become
\eqn\ck{ \eqalign{G_{IJ} F^I_{m\th} &= {1\over 2} X_I f^{-1}\p_m f~,\cr
F^I_{mn}  &= f X^I (d\omega)_{mn}+ \Theta^I_{mn}~.}}
Using \aib\ we find that the second
line of \ci\ is satisfied provided
\eqn\cl{ \p_i X_I \Theta^I_{mn}=0~,}
which in turn requires that $\Theta^I$ has the  structure
\eqn\cm{ \Theta^I_{mn} = X^I k_{mn}~.}

So to summarize, enhancement of supersymmetry implies the two
conditions \cj\ and \cm.

\subsec{Charges}

The power of the attractor mechanism is that it fixes moduli in
terms of the charges carried by the black hole, whether of the
conserved or dipole variety.   We therefore need to give formulas
for the charges, which we do in this subsection.

Let $V$ be some bounded region in the base ${\cal B}$.  We define
the electric charge in $V$ as
\eqn\daa{ Q_I(V) ={1 \over 2\pi^2}  \int_{\p V} \! dS\, f^{-1}
{n}^mE_{ mI}~,}
where ${n}$ is the outward pointing unit normal vector. The
conserved electric charge measured at infinity is obtained by
taking $V= {\cal B}$.  In general the value of the charge as we
have defined it  depends nontrivially on $V$;  indeed, from \cfa\
we have
\eqn\dab{ Q_I(V_1) - Q_I(V_2) = -{1 \over 16\pi^2} \int_{V_1 -V_2}
\!d^4x \, \sqrt{h}\, C_{IJK} \Theta^J \cdot \Theta^K~.}
Since it is the dressed field $X^I E_{mI}$ that appears in the
flow equation \ch, it is  natural to define also
\eqn\dc{Z_e(V)  = {1 \over 2\pi^2}  \int_{\p V} \! dS\, f^{-1} X^I
{n}^mE_{ mI}~,}
which obeys
\eqn\dd{  Z_e(V_1) - Z_e(V_2)= \int_{V_1 - V_2} \!d^4x \,
\sqrt{h}\, \left\{ {1 \over 2\pi^2 }f^{-1} G_{IJ} \nabla^m X^I
\nabla_m X^J -{1 \over 16\pi^2}  C_{IJK} X^I\Theta^J \cdot
\Theta^K\right\}~.}
$Z_e$ is the electric charge corresponding to the graviphoton. As
measured at infinity, it  is also the central charge appearing in
the BPS mass formula:
\eqn\de{ M =  Z_e({\cal B})~.}

We next turn to the definition of the {\it dipole charges} $q^I$,
which are defined as integrals of $\Theta^I$ over certain
noncontractible 2-spheres in ${\cal B}$:
\eqn\df{ q^I= - {1 \over 2\pi} \int_{S^2} \Theta^I~.}
These 2-spheres can arise in either of two ways.   First, the base
${\cal B}$ may be a smooth four manifold supporting such
noncontractible spheres.   Alternatively, ${\cal B}$ could be
topologically trivial, such as flat $\IR^4$. In this case,
$\omega$ and $\Theta^I$, viewed as differential forms on ${\cal
B}$, may have singularities even though the full five-dimensional
geometry is smooth.   If these singularities lie along a closed
curve, as is the case for the black ring solutions, then there
will be noncontractible 2-spheres which surround the curve.   In
either case, we define the dipole charges as in \df. In analogy
with \dc\ we also define
\eqn\dg{ Z_m = - {1 \over 2\pi} \int_{S^2} X_I\Theta^I~.}
One interpretation of $Z_m$ is that if we take the singular curve
described above to be an infinite straight line, then our solution
will describe a magnetic string whose BPS mass formula is governed
by $Z_m(\infty)$.   From the M-theory point of view, such magnetic
strings can be realized as M5-branes wrapping 4-cycles of the
CY$_3$.

In many considerations the string-like charges appear on more or
less equal footing with the more familiar electric charges. For
example,  the $Z_e$ and the $Z_m$ at infinity both appear  in the
supersymmetry algebra when we allow for extended string solutions.
%
%\eqn\susyalg{
%\{ Q_{A\alpha}, Q_{B\beta} \} = ({\cal C}\Gamma^M)_{\alpha\beta}  P_M \Omega_{AB}
%+ {\cal C}_{\alpha\beta} Z_{eAB} +  ({\cal C}\Gamma_M)_{\alpha\beta} Z^M_{mAB}
%}
%where $A,B=1,2$.
Similarly, we will see that there are attractors controlled by the
dipole charges, which are quite similar to the usual attractors
dominated by point-like charges.

\subsec{Attractor Flows With $\Theta^I=0$} We now consider the
special case in which the dipole field strengths vanish,
$\Theta^I=0$.  This special case includes the BMPV black hole
\BMPV, and more generally, multi-centered versions of these.

For this case there is a monotonic flow of the central charge
$Z_e$.  In particular, from \dd, we see that if $V_2$ is contained
within $V_1$ then
\eqn\ea{ Z_e(V_1) - Z_e(V_2)= {1 \over 2\pi^2} \int_{V_1 -V_2}
\!d^4x \, \sqrt{h}\, f^{-1} G_{IJ} \nabla^m X^I \nabla_m X^J ~
\geq 0 ~.}
From \dab\ we also see that $Q_I$ is independent of $V$.\foot{We
stress that  this independence is only true provided no
singularities pass into or out of $V$ as we deform it.}

The typical situation is for there to be isolated pointlike
singularities on ${\cal B}$, whose locations can intuitively be
thought of as specifying the locations of branes.  Take $V$ to
enclose a single singularity at $P$. The central charge decreases
monotonically as the singularity is approached and, if the
singularity is not too severe, it will reach a finite value in the
limit. Furthermore, if there is a smooth black hole horizon at $P$
then $f(P)=0$, and the factor $\sqrt{h} f^{-1}$ diverges.
 Finiteness of $Z_e(P)$ then  typically forces $\nabla_m X^I|_P=0$,
  and so we are in the
situation where supersymmetry is enhanced, as discussed in section
4.2. This is the attractor we want to study.

The essence of the attractor mechanism is that we can work out the
values taken by the $X^I$ at $P$ in terms of the charges $Q_I$. To
see this, we start contracting $V$ around the point $P$. For
sufficiently small $V$ we have, from \cd, \ck, and \daa,
\eqn\ec{ Q_I = X_I \left(-{1 \over 4\pi^2} \int_{\p V} \! dS \,
n^m \p_m f^{-1}\right) ~.}
The term in the bracket is just a constant of proportionality
determined by the condition of unit volume for the CY$_3$.
This gives the fixed point values
\eqn\ed{X_I ={Q_I \over ({1 \over 6} C^{JKL} Q_J Q_K Q_L)^{1/3}
}~.}

An equivalent way of stating this is that we can find the fixed
values of the moduli by extremizing the central charge $Z_e$.  In
particular, at $P$
\eqn\ee{ Z_e = X^I(P) Q_I~.}
Extremizing the central charge with respect to the fixed moduli
means that we impose
\eqn\ef{ \p_i Z_e =0~.}
The $Q_I$ are held fixed, so the equation reads $\p_i X^I(P)
Q_I=0$.  But this equation implies that $Q_I$ is proportional to
$X_I$, which then leads to \ed\ in the same way that  \ec\ led to
\ed.

We see that the moduli take values at $P$  determined by the
charges at $P$.  For a single singularity these charges are the
same as the conserved charges measured at infinity, and $Z_e$ is
the central charge appearing in the the BPS mass formula.  More
generally, for  multiple singularities  the charge at infinity is
a sum of contributions from each singularity, and so it is not
possible to read off the various fixed moduli directly from the
charge at infinity.

As an example, let's consider the BMPV black hole. We take the
compactification manifold to be $T^6$, as in \aj-\al.  In this
case the base metric is flat
\eqn\eg{ h_{mn} dx^M dx^N = dr^2 + r^2( d\theta^2 + \sin^2 \theta
d\psi^2 + \cos^2 \theta d\phi^2) ~.}
The five dimensional metric is given by \bb\ with
\eqn\eh{\eqalign{f^{-3} &= {1 \over 6} C^{IJK} H_I H_J H_K~, \quad
H_I = 1 +{Q_I \over r^2}~,  \cr \omega & = -{4 G_5 \over \pi} J
(\cos^2 \theta d\phi + \sin^2 \theta d\psi)~.}}
The solution carries angular momentum $J_\psi = J_\phi =J$.
The gauge fields and moduli are
\eqn\eh{\eqalign{ E_{rI} &= f{Q_I \over r^3}~, \cr   X_I &= {H_I
\over (H_1 H_2 H_3)^{1/3}}~.}}
The horizon is at $r=0$.  At the horizon the moduli take values
\eqn\ei{ X_I(r=0) = {Q_I \over (Q_1 Q_2 Q_3)^{1/3}}~,}
in agreement with \ed. We define the attractor function $Z_e$ as
in  \dc\ with $V$ taken
to be a 3-sphere of radius $r$.  $Z_e$ then takes the form
\eqn\ej{ Z_e(r) = X^I Q_I = {(H_1 H_2 H_3)^{1/3}} H_I^{-1}Q_I~,}
and obeys ${d Z_e \over dr} \geq 0$ in agreement with \ea. $Z_e$
interpolates between the following two values:
\eqn\ek{ Z_e(r=0) = 3(Q_1 Q_2 Q_3)^{1/3}, \quad Z_e(r=\infty) =
Q_1 + Q_2 +Q_3~.}
The value at the origin determines the radius of curvature of a
near horizon AdS$_2 \times S^3$ geometry, while the value at
infinity gives the mass according to \de.

\subsec{Attractor Flows With $\Theta^I \neq 0$}

We now consider the case of nonzero dipole field strengths
$\Theta^I$, and in particular, nonzero dipole charges $q^I$.   In
this case the moduli are {\it not} fixed in terms of the charges
$Q_I$, and $Z_e$ does {\it not} behave monotonically.  Instead,
the attractor flow is governed by the dipole charges $q^I$ and the
attractor function is $Z_m$.  We first discuss this in general
terms, and then illustrate using the example of the black ring
solution.

We assume that there is a noncontractible $S^2$ in order to define
the dipole charges $q^I$.  The $S^2$ is taken to surround a closed
curve on the base ${\cal B}$.  We further assume that there are
enough nonzero dipole charges so that $C_{IJK}q^I q^J q^K \neq 0$.
Finally, we assume enhanced supersymmetry as we approach the
curve. Under these assumption the near horizon geometry becomes
$AdS_3\times S^2$ (rather than $AdS_2\times S^3$ for the case with
no dipole charges). Typically,  if the above conditions are not
met then the moduli will not be stabilized and there will not be a
regular horizon either. We should emphasize that this doesn't
necessarily imply that the geometry is singular, just that there
is not a regular black hole horizon.  There might instead be a
smooth horizon-free geometry.  But in such a case there is no
expectation that the attractor mechanism will be operative.

Let us now return to the conditions for enhanced supersymmetry.
We first observe that \cm\ and \df\ imply
\eqn\ga{q^I  = \left({1 \over 2\pi} \int_{S^2} k\right) X^I~.}

Whenever $\int_{S^2} k \neq 0$, there is enough information to
determine the near horizon values of $X^I$ in terms of the dipole
charges $q^I$.  In particular, the solution is
\eqn\gc{ X^I = {q^I \over ({1 \over 6} C_{IJK} q^I q^J
q^K)^{1/3}}~.}

As before, an equivalent way of arriving at \gc\ is to consider
$Z_m = X_I q^I$, and demand $\p_i Z_m =0$.   This leads to \gc\
since the vanishing of $\p_i Z_m = \p_i X_I q^I$ implies that
$X^I$ is proportional to $q^I$,  which is the content of  \ga.

 We now illustrate the above with the black ring example \EEMR.
 The compactification manifold is $T^6$ as in \aj-\al. The base metric
  is flat,
\eqn\gcm{\eqalign{  h_{mn} dx^m dx^n & = dr^2 + r^2 (d\theta^2 +
\sin^2 \theta  + \cos^2 \theta d\phi^2) \cr
 & = {R^2 \over (x-y)^2} \left[ {dy^2 \over y^2 -1} +(y^2-1)
 d\psi^2 + {dx^2 \over 1-x^2} +(1-x^2) d\phi^2 \right]~,}}
 where in the second coordinate system $x$ and $y$ have range: $-1\leq
 x \leq 1$, $-\infty \leq y \leq -1$.    The solution has
\eqn\gd{\eqalign{ X_I &= {H_I \over ( H_1 H_2 H_3)^{1/3}} ~,\cr H_I
&= 1 + {Q_I - \half C_{IJK} q^J q^K \over 2 R^2} (x-y) - {C_{IJK}
q^J q^K \over 8 R^2} (x^2 -y^2)~.}}
The field strengths take the form \bc\ with the dipole field
strength given by
\eqn\ge{ \Theta^I =- \half q^I ( dy \wedge d\psi + dx \wedge
d\phi)~.}
We also have
\eqn\gea{ f^{-3} = {1 \over 6} C^{IJK}H_I H_J H_K~.}
 Finally, the
1-form $\omega$ is \eqn\geo{\eqalign{ \omega_\psi & = - {1\over
8R^2} (1-x^2) \left[ Q_I q^I - {1\over 6} C_{IJK} q^I q^J q^K
(3+x+y)\right]~,\cr \omega_\phi &=  {1\over 2} (q^1 + q^2 + q^3)
(1+y) + \omega_\psi~. }}

The horizon of the black ring is at $y=-\infty$ where the $H_I$
diverge.   $\psi$ is is the angular coordinate parameterizing the
location along the ring.  The noncontractible 2-spheres
surrounding the ring are parameterized by $x$ and $\phi$. Given
\ge\ it is indeed easy to see that integration on the 2-spheres
gives
\eqn\gf{ q^I =  - {1 \over 2\pi} \int_{S^2} \Theta^I~,}
as in \df.  Also, in the limit $y\rightarrow -\infty$ we see from
\gd\ that the moduli take on values in agreement with \gc.

We now discuss the behavior of the attractor functions $Z_e$ and
$Z_m$ in the full black ring geometry.  Neither function is
monotonic; rather, as the ring is approached from infinity there
is a sort of crossover, with $Z_e$ decreasing for large radius,
and $Z_m$ decreasing for small radius.  We can think of this in
terms of an RG flow, where we interpolate between a UV CFT
controlled by the electric charges $Q_I$, and an IR CFT controlled
by the dipole charges $q^I$.

$Z_e$ is given by \dc, where it natural to take $V$ to be an $S^3$
of radius $r$.  A little manipulation yields
\eqn\gg{ Z_e(r)=- {3 \over 4\pi^2} \int\! dS\, \p_r f^{-1}~.}
The explicit formula for $f$ is
\eqn\gh{ f=(H_1 H_2 H_3)^{-{1\over 3}}~,}
with
\eqn\gi{\eqalign{ H_I &= 1 + {Q_I - \half C_{IJK} q^J q^K \over
\Sigma} + \half C_{IJK} q^J q^K {r^2 \over \Sigma^2}~,\cr \Sigma &=
\sqrt{(r^2 -R^2)^2 + 4R^2 r^2 \cos^2 \theta}~.}}
For large $r$ the $Q_I$ dominate and $f$ behaves as in the BMPV
case \eh.  But near the ring $Z_e$ behaves as
\eqn\gj{ Z_e(r) \sim 6 {(q^1 q^2 q^3)^{2/3} R^6 \over (r^2
-R^2)^3} \quad {\rm as}\quad r\rightarrow R~.}
So in the full geometry $Z_e$ is neither monotonic nor bounded.

To define $Z_m$ we instead work in the $x-y$ coordinates and
integrate over 2-spheres of fixed $y$ and $\psi$.  From \dg\ and
\ge\ this gives
\eqn\gk{ Z_m(y) = {1 \over 2} \int_{-1}^1 \! dx\, q^I X_I~. }
At the horizon $y\rightarrow -\infty$, and $Z_m$ stabilizes at
$Z_m = q^I X_I$ with $X_I$ given by \gc. In more detail, the near
horizon behavior is
\eqn\gl{ Z_m = 3(q^1 q^2 q^3)^{1/3} +{2 \over 3} {(q^1 Q_1 -q^2
Q_2)^2 +(q^2 Q_2 -q^3 Q_3)^2 +(q^3 Q_3 -q^1 Q_1)^2 \over (q^1 q^2
q^3)^{5/3}}y^{-2}  + O(y^{-3})~.}
So sufficiently near the ring $Z_m$ is decreasing as it approaches
its fixed values.  However, analysis of the integral \gk\ reveals
that farther from the ring $Z_m$ is not monotonic.

While neither $Z_e$ nor $Z_m$ is monotonic throughout the full
geometry, one might ask whether some other combination of charges
and moduli does better in this regard.  Our analysis does not
reveal any obvious candidate, and in particular there is no natural
family of  surfaces on which to define such an expression.

\subsec{Extremization Principles}
As we have seen, the near horizon moduli are fixed by one of the
two attractor functions:  $Z_e = X^I Q_I$ or $Z_m= X_I q^I$.   For
$\Theta^I =0$,  the near moduli are fixed by extremizing $Z_e$;
{\it i.e.} solving  $\p_i Z_e=0$.   On the other hand, for sufficiently
many nonzero dipole charges the moduli are instead found by
extremizing $Z_m$.   To be clear, we note that when we talk of
extreming $Z_e$ and $Z_m$, we mean extremization with respect to
the moduli $X^I$ while holding the charges $Q_I$ and dipole
charges $q^I$ fixed ($G_5$ is fixed throughout, {\it via} our choice of
units).

It is also natural to rephrase the extremization procedure in
terms of physical quantities measured at infinity.  First consider
the case of $\Theta^I=0$.  The general BPS mass formula is written
in  \de\ (this is valid for the general case with nonzero
$\Theta^I$.)  If we allow for arbitrary $X^I$ at infinity, then we
see that the BPS mass depends nontrivially on these $X^I$.  We
also see that the value of the $X^I$ which extremizes the mass
(while holding the charges fixed) are the same values as appear in
the near horizon region of enhanced susy.   Therefore, the near
horizon moduli can be determined purely from considerations of the
BPS mass measured at infinity.  Recalling that $Z_e$ is monotonic,
we also learn that if we choose moduli at infinity so as to
minimize $Z_e$, then it must be the case that $Z_e$ is constant
throughout the flow to the horizon.  This also implies that $X^I$
are similarly constant throughout the flow in this case.

Now turn to the case of nonzero dipole charges such that the
moduli are fixed by $Z_m$.  Is it again possible to rephrase this
in terms of extremizing some quantities of direct physical
relevance?  We will present two complimentary extremization
principles based on the explicit black ring solutions.

First, we recall that in five dimensions there are two independent
angular momenta, which we are calling $J_\psi$ and $J_\phi$.  For
the black ring the difference of these is
\eqn\ya{ J_\psi - J_\phi = R^2 {\bar X}_I q^I = R^2 Z_m~,}
where $R^2$ is the ring radius, and the ${\bar X}_I$ refer to the
moduli at infinity. This combination of angular momenta can be
interpreted as the intrinsic angular momentum of the ring (not
including the surrounding field). Alternatively, it can be identified
with the effective level of the dual CFT \foot{Heuristically, this is the
momentum flowing along the string; more precisely, the effective
level $h_{\rm eff}$ is the eigenvalue of $L_0 -{\bar L}_0$ when
the supersymmetric sector of the $(4,0)$ dual CFT is in its
NS-sector ground state.}. Here
we find that that extremizing $Z_m$ is the same as extremizing
$J_\psi - J_\phi$ while holding $q^I$ and $R$ fixed.  It is not
clear whether a version of this principle is true in general.  In
particular, for a well defined formulation one needs to give
meaning to the ring radius $R$ for a general solution, not
necessarily of the black ring form.  We leave this as an open
question.

Another interesting extremization principle emerges from the black ring
entropy formula \refs{\BenaDE,\EEMR}
\eqn\yb{ S=2\pi \sqrt{J_4}~, \quad J_4 = J_4(Q_I ,q^I,
J_\psi-J_\phi)~.}
Here $J_4$ is the quartic $E_{7(7)}$ invariant; see
\refs{\bkmicro,\cgms} for the explicit formula. Therefore, we can
obtain the fixed moduli by extremizing the entropy while holding
fixed $Q_I$, $q^I$, and $R$. This suggests a thermodynamic
interpretation of the attractor mechanism. Again, a general
version of this requires formulating a  general notion of $R$.

\newsec{Discussion}

We have obtained results on attractor flows in the context of
general BPS solutions in five dimensions, generalizing earlier
work on this subject.  In particular, we saw that nonzero dipole
charges, which are natural quantities to define in the context of
real special geometry, lead to a new type of attractor flow
governed by the attractor function $Z_m$.  For this class of
flows, which includes the recently discovered BPS black rings, the
near horizon moduli take values governed by the dipole charges,
rather than by the conserved electric charges measured at
infinity.

There are several issues worthy of better understanding.  First,
while $Z_e$ is governed by the flow equation \dd\ we did not find
any analogous equation for $Z_m$.  Technically, this was because
the $\hat{t}$ index appearing in the projection equation \ca\
singled out the electric field as playing a special role.  On the
other hand, for the black rings we did observe that $Z_m$ was
decreasing as we got near the ring, and it would be nice to have a
derivation of this from general principles.

A better understanding of the interplay between $Z_e$ and $Z_m$
might also shed light on general aspects of RG flows. For example,
the standard near-horizon decoupling limit of the black ring
solutions yields a solution describing an RG flow from a UV CFT
with central charge set by $Z_e$ to an IR CFT with central charge
set by $Z_m$. From the explicit solution it always turns out that
$c_{IR} \leq c_{UV}$.  It would be interesting to establish that
this is always the case.

Finally, it remains to be determined what, precisely, is the
physical principle responsible for the attractor mechanism. There
is some evidence that the attractor equations follows from the
extremization of thermodynamic potentials such as the entropy, but
the general formulation of such a physical principle is an open
problem.

%\vskip 3.0cm
\bigskip
\noindent {\bf Acknowledgments:} \medskip \noindent We would like
to thank  Iosif Bena, Henriette Elvang, Jim Liu, and
Erik Verlinde  for discussions.
FL thanks UCLA's IPAM for hospitality when this work was initiated.
The work of PK is supported in part by the NSF grant PHY-00-99590.
The work of FL is supported in part by the DoE.

\listrefs
\end

For concreteness {\bf (may want to relax)} let us take the base
metric to be flat: $h_{mn} dx^m dx^n = dr^2 + r^2 d\Omega_3^2$. We
define the charge $Q_I(r)$  as
\eqn\da{ Q_I(r)  = {1 \over 2\pi^2} \int_{S^3} \! dS\, f^{-1} E_{r
I}~.}
The conserved electric charge measured at infinity is
$Q_I(r=\infty)$.    In general $Q_I$ is a nontrivial function of
$r$; indeed, from \cfa\ we have
\eqn\db{{d Q_I \over dr} = -{1 \over 4} r^3 C_{IJK} \Theta^J \cdot
\Theta^K~.}
Given \ch\ it is also natural to define
\eqn\dc{ Z_e = X^I Q_I~,}
which obeys
\eqn\dd{ {d Z_e \over dr} = {r^3 \over f} G_{IJ} \nabla^m X^I
\nabla_m X^J-{1 \over 4} r^3 C_{IJK}X^I \Theta^J \cdot \Theta^K~.}
$Z_e$ is the electric charge corresponding to the graviphoton.  It
is also the central charge appearing in the BPS mass formula:
\eqn\de{ M = {\pi \over 4 G_5} Z_e(\infty)~.}
